# Comparison of Random Waypoint & Random Walk Mobility Model under DSR, AODV & DSDV MANET Routing Protocols

Muhammad Zaheer Aslam
Student of MSCS, Gomal University
DIKhan.
zaheer.aslam@yahoo.com

Dr. Abdur Rashid
Associate Professor, Gomal University
DIKhan.
rashidkh08@yahoo.com

*Abstract-* **Mobile Adhoc Network is a kind of wireless ad hoc network where nodes are connected wirelessly and the network is self configuring. MANET may work in a standalone manner or may be a part of another network. In this paper we have compared Random Walk Mobility Model and Random Waypoint Mobility Model over two reactive routing protocols Dynamic Source Routing (DSR) and Adhoc On-Demand Distance Vector Routing (AODV) protocol and one Proactive routing protocol Distance Sequenced Distance Vector Routing (DSDV) Our analysis showed that DSR, AODV & DSDV under Random Walk and Random Way Point Mobility models have similar results for similar inputs however as the pause time increases so does the difference in performance rises. They show that their motion, direction, angle of direction, speed is same under both mobility models. We have made their analysis on packet delivery ratio, throughput and routing overhead. We have tested them with different criteria like different number of nodes, speed and different maximum number of connections.**

1.        INTRODUCTION

This is the age of Wireless communication systems. These days there is a need of rapid deployment of independent mobile users, for example establishing survivable, dynamic communication for emergency operations, disaster management and military networks, crime management networks etc. These all types of networks are actually based on the mobile adhoc networks. Such networks don't have a central control but they are decentralized networks. MANET is autonomous collection of mobile nodes that communicate over limited bandwidth and energy constraints. These mobile nodes are in motion so the topology of the entire network changes rapidly and unpredictably over time. All network is managed by the network nodes themselves, as there is no special device or router involved, every nodes itself work as a router to forward the traffic.
Routing in these types of networks is a main issue as there is no fixed infrastructure and paths get changed due to rapid movement of nodes, so routing is a main area where research needs to be done. There is need of some special routing protocols for these types of networks which can automatically recognize the topology changes and which can limit the extra overhead of control messages before data transfer as these networks has low bandwidth.
MANET Routing protocols are divided into two categories: Proactive and Reactive. Proactive routing protocols are table-driven protocols and they always maintain current up-to-date routing information by sending control messages periodically between the hosts which update their routing tables. The proactive routing protocols use link-state routing algorithms which frequently flood the link information about its neighbors. [2] Reactive or on-demand routing protocols create routes when they are needed by the source host and these routes are maintained while they are needed. Such protocols use distance-vector routing algorithms. [1]
Our goal is to carry out a systematic performance study of three routing protocols for ad hoc networks namely Ad hoc On Demand Distance Vector (AODV) Routing protocol, Dynamic Source Routing (DSR) protocol and Distance Sequenced Distance Routing protocol under Random Walk Mobility Model and Random Waypoint Mobility Model.

The rest of the paper is organized as follows: Section 2 gives a brief description of the routing protocols used for performance comparison. Section 3 gives description of mobility models used in the paper. Section 4 gives study of previous related work done. In Section 5 we present the setup of the Simulation Environment. Section 6 gives the Results and Analysis of the simulation done, Section 7 is conclusion while at last section 8 provide the references.

## 2. DESCRIPTION OF PROTOCOLS

### 2.1 DSDV

Destination Sequenced Distance Vector protocol belongs to the class of pro-active routing protocols. This protocol is based on the classical Bellman-Ford routing algorithm [4] to apply to mobile ad hoc networks. DSDV also has the feature of the distance- vector protocol [5] in that each node holds a routing table including the next-hop information for each possible destination. Each entry has a sequence number. If a new entry is obtained, the protocol prefers to select the entry having the largest sequence number. If their sequence number is the same, the protocol selects the metric with the lowest value. Routing information is transmitted by broadcast. Updates have to be transmitted periodically or immediately when any significant topology change is available. Sequence numbers are assigned by destination, means the destination gives a sort of default even sequence number, and the emitter has to send out the next update with this number. Packets are transmitted between the stations of the network by using routing tables which are stored at each station of the network. Each routing table, at each of the stations, lists all available destinations, and the number of hops to each. Each route table entry is tagged with a sequence number which is originated by the destination station. To maintain the consistency of routing tables in a dynamically topology, each station periodically transmits updates, and transmits updates immediately when significant new information is available. Routing information is advertised by broadcasting or multicasting the packets which are transmitted periodically and incrementally as topological changes are detected - for instance, when stations move within the network. Data is also kept about the length of time between arrival of the first and the arrival of best route for each destination. Based on this data, a decision may be made to delay advertising routes which are about to change soon, thus damping fluctuations of the route tables.

### 2.2 DSR

The Dynamic Source Routing (DSR) protocol is an on-demand routing protocol based on source routing. In the source routing technique, a sender determines the exact sequence of nodes through broadcasted route request Message. When route is found then route reply is made containing the route to destination. The list of intermediate nodes for routing is explicitly contained in the packet's header. In DSR, every mobile node in the network needs to maintain a *route cache* where it caches source routes that it has learned. When a host wants to send a packet to some other host, it first checks its route cache for a source route to the destination. In the case a route is found, the sender uses this route to propagate the packet. Otherwise the source node initiates the route discovery process. Route discovery and route maintenance are the two major parts of the DSR protocol.

### 2.3 AODV

This protocol performs Route Discovery using control messages route request (RREQ) and route reply (RREP) whenever a node wishes to send packets to destination. To control network wide broadcasts of RREQs, the source node uses an *expanding* ring search technique. The forward path sets up an intermediate node in its route table with a lifetime association RREP. When either destination or intermediate node using moves, a route error (RERR) is sent to the affected source node. When source node receives the (RERR), it can reinitiate route if the route is still needed. Neighborhood information is obtained from broadcast Hello packet. As AODV protocol is a flat routing protocol it does not need any central administrative system to handle the routing process. AODV tends to reduce the control traffic messages overhead at the cost of increased latency in finding new routes. The AODV has great advantage in having less overhead over simple protocols which need to keep the entire route from the source host to the destination host in their messages. The RREQ and RREP messages, which are responsible for the route discovery, do not increase significantly the overhead from these control messages. AODV reacts relatively quickly to the topological changes in the network and updating only the hosts that may be affected by the change, using the RRER message. The Hello messages, which are responsible for the route maintenance, are also limited so that they do not create unnecessary overhead in the network. The AODV protocol is a loop free and avoids the counting to infinity problem, which were typical to the classical distance vector routing protocols, by the usage of the sequence numbers. [3]

## 3. MOBILITY MODELS

There are two types of mobility models used in the simulation of networks: traces and synthetic models [8,9]. Traces are those mobility patterns that are observed in real life systems. They provide accurate information when they involve a large number of nodes and an appropriately long observation time. However, new network environments like ad hoc networks are not easily modeled if traces have not yet been created. In this type of situation it is necessary to use synthetic models. Synthetic models attempt to realistically represent the behaviors of MNs without the use of traces. Different synthetic entity mobility models for ad hoc networks are [9]
1. Random Walk Mobility Model (including its many derivatives): A simple mobility model based on random directions and speeds.
2. Random Waypoint Mobility Model: A model that includes pause times between changes in destination and speed.
3. Random Direction Mobility Model: A model that forces MNs to travel to the edge of the simulation area before changing direction and speed.
4. A Boundless Simulation Area Mobility Model: A model that converts a 2D rectangular simulation area into a torus-shaped simulation area.
5. Gauss-Markov Mobility Model: A model that uses one tuning parameter to vary the degree of randomness in the mobility pattern.
6. A Probabilistic Version of the Random Walk Mobility Model: A model that utilizes a set of probabilities to determine the next position of an MN.
7. City Section Mobility Model: A simulation area that represents streets within a city.
In this paper we are analyzing the first two models.

### 3.1) The Random Walk Mobility Model
It was first described mathematically by Einstein in 1926 [9]. Since many entities in nature move in extremely unpredictable ways, the Random Walk Mobility Model was developed to mimic this erratic movement. In this mobility model, an MN moves from its current location to a new location by randomly choosing a direction and speed in which to travel. The new speed and direction are both chosen from pre-defined ranges, [*speedmin*; *speedmax*] and [0;2π] respectively. Each movement in the Random Walk Mobility Model occurs in either a constant time interval *t* or a constant distance traveled *d*, at the end of which a new direction and speed are calculated. If an MN which moves according to this model reaches a simulation boundary, it "bounces" off the simulation border with an angle determined by the incoming direction. The MN then continues along this new path. Many derivatives of the Random Walk Mobility Model have been developed including the 1-D, 2-D, 3-D, and d-D walks.

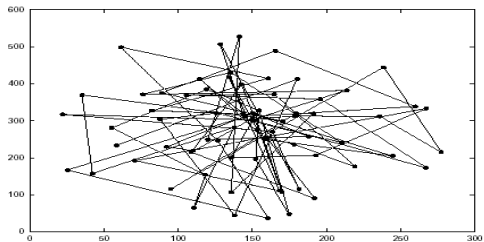

Figure 1: Traveling pattern of an MN using the 2-D Random Walk Mobility Model (time).

### 3.2) Random Waypoint Mobility Model
It includes pause times between changes in direction and/or speed. An MN begins by staying in one location for a certain period of time (i.e., a pause time). Once this time expires, the MN chooses a random destination in the simulation area and a speed that is uniformly distributed between *[minspeed,maxspeed]*. The MN then travels toward the newly chosen destination at the selected speed. Upon arrival, the MN pauses for a specified time period before starting the process again. Figure 2 shows an example traveling pattern of an MN using the Random Waypoint Mobility Model starting at a randomly chosen point or position (133, 180); the speed of the MN in the figure is uniformly chosen between 0 and 10 m/s. We note that the movement pattern of an MN using the Random Waypoint Mobility Model is similar to the Random Walk Mobility Model if pause time is zero and *[minspeed, maxspeed] = [speedmin, speedmax]*. [9]

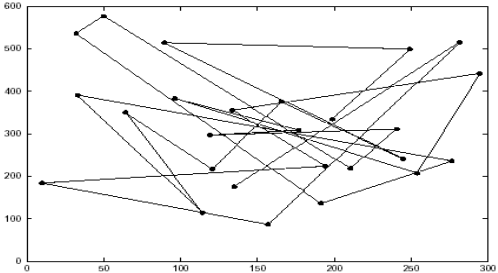

Figure 2: Traveling pattern of an MN using the Random Waypoint Mobility Model.

## 4. RELATED WORK

In [6] four different routing protocols AODV, TORA, DSDV and DSR are compared. Simulation showed that DSR generates less routing load compared to AODV. AODV suffers from end to end delay while TORA has very high routing overhead. The better performance of DSR is due to efficient use of cache and maintains multiple routes to the destinations.

Perkins in [10] showed the performance of DSR and AODV. Since both AODV and DSR use on demand route discovery but they have different routing mechanics. The authors Show that delay and throughput of DSR outperforms AODV when the numbers of nodes are smaller. AODV outperforms DSR when the number of nodes is very large. The authors do show that DSR consistently generate less routing load than AODV.

[7] has comparison of Link State, AODV and DSR protocols for two different traffic classes, in a selected environment. It is claimed that AODV and DSR perform well when the network load is moderate and if the traffic load is heavy then simple Link State outperforms the reactive protocols.

Performance comparison of AODV and DSR is done in [11]. The authors says that the AODV outperforms DSR in normal situation but in the constrained situation DSR out performs AODV, where the degradation is as severe as 30% in AODV whereas DSR degrades marginally as 10%.

## 5. SIMULATION ENVIRONMENT

A lot of simulations have been carried out for analysis of routing protocols. We have used ns-2 [13] for our analysis using cygwin running on windows operating system. Our simulation time is 200 sec and area chosen is 670X670. We have used 10 and 50 number of nodes for testing generating packet size of 512 bytes, with maximum connections of 20% and 60%. We have used one packet per second traffic generation rate. For Random Walk mobility Model we have used pause time zero where as for Random Way point Mobility model we have used pause time of 25 seconds. We have applied maximum speed of 10 and 50 for node movement.

We have used a tool setdest [1] which comes with network simulator 2 for generation of scenario files. In our simulation we have generated CBR traffic. For CBR traffic we have used the builtin tcl program of ns2 called cbrgen.tcl [12].

## 6. RESULTS AND ANALYSIS

Our results show clearly that Random Walk and Random Waypoint Mobility Model both are same models actually. There motion is same but one difference is pause time which is zero in Random Walk mobility model. As we increase the pause time in Random Waypoint Mobility model, the motion decreases and the path linkage break also get decreases and so does the performance variance starts with compared to Random Walk Mobility Model.

**1) Packet Delivery Ratio**
According to David Oliver Jörg (2003), packet delivery ratio is calculated by dividing the number of packets received by the destination through the number of packets originated by the application layer of the source (i.e. Constant Bit Rate (CBR)). It specifies the packet loss rate, which limits the maximum throughput of the network. The better the delivery ratio, the more complete and correct is the routing protocol.

Comparing the results of Fig 3(a) to Fig 3(b) and of Fig 3(c) to Fig 3(d), we conclude that all protocols react same under Random Walk Mobility Model and random Waypoint Mobility model. Apart from this we also conclude that reactive protocols have greater packet delivery ratio compare to proactive protocol.

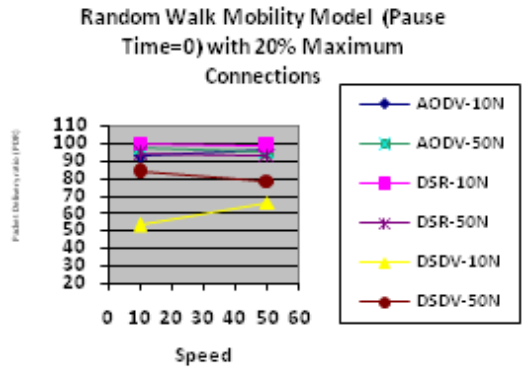

Fig 3(a)

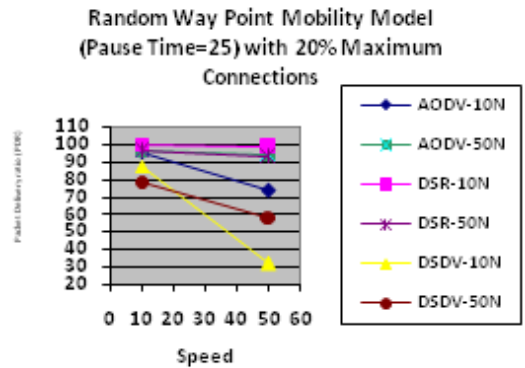

Fig 3(b)

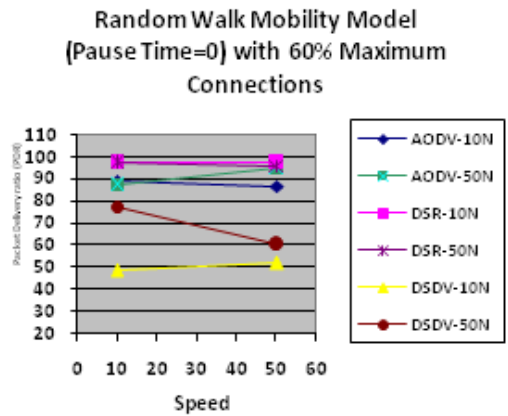

Fig 3(c)

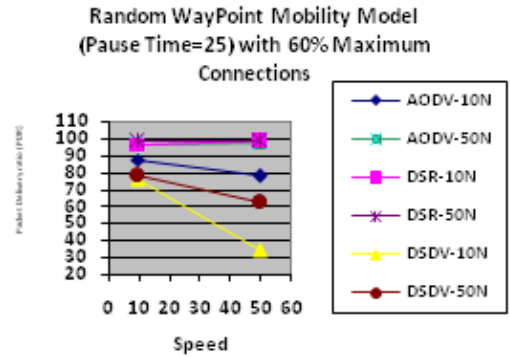

Fig 3(d)

**2) Normalized Routing Overhead**

Normalized routing overhead is the total number of routing packets divided by total number of delivered data packets (A. Al-Maashri and M. Ould-Khaoua, 2006). In the context of this project, the average number of routing packets required to deliver a single data packet is analyzed. This metric provides an indication of the extra bandwidth consumed by overhead to deliver data traffic. It is crucial as the size of routing packets may vary.

Comparing the results of Fig 4(a) to Fig 4(b) and of Fig 4(c) to Fig 4(d), we conclude that all the protocols react same under Random Walk Mobility Model and random Waypoint Mobility model. It is also clear that proactive protocol has more routing overhead compared to reactive protocols.

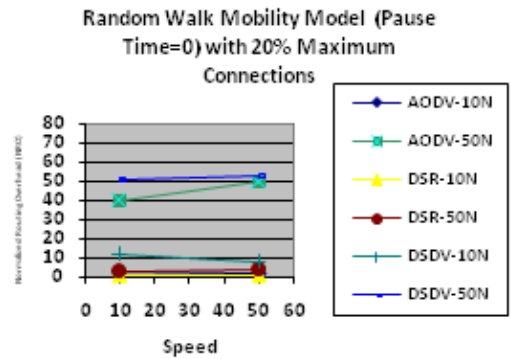

Fig 4(a)

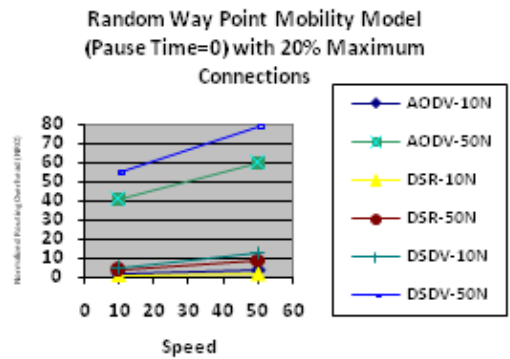

Fig 4(b)

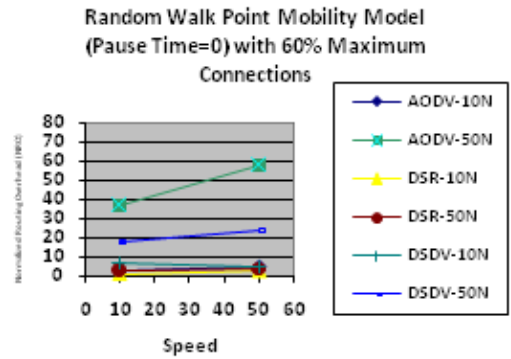

Fig 4(c)

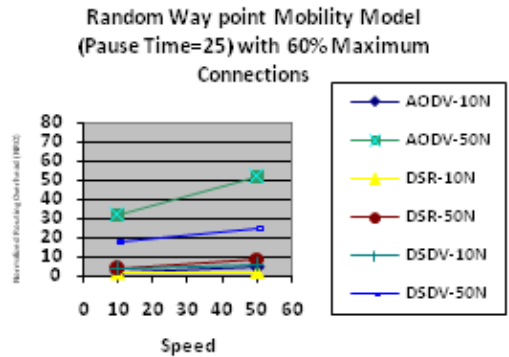

Fig 4(d)

**3) Throughput**
The throughput (messages/second) is the total number of delivered data packets divided by the total duration of simulation time or the throughput of each of the routing protocol in terms of number of messages delivered per one second is evaluated.

Comparing Fig 5(a) to Fig 5(b) and Fig 5(c) to Fig 5(d), we say that all three protocols react equally same under Random Walk Mobility Model and random Waypoint Mobility model. We also conclude that DSDV being a proactive protocol gives low throughput compared to others. Similarly protocols increase with increasing number of nodes.

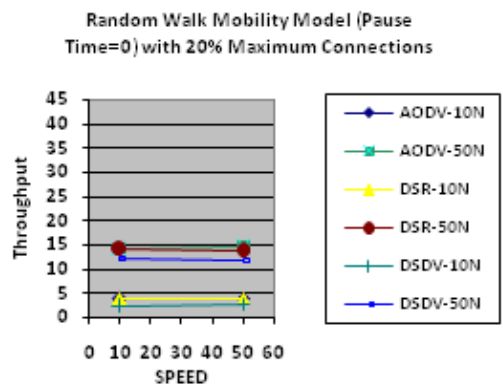

Fig 5(a)

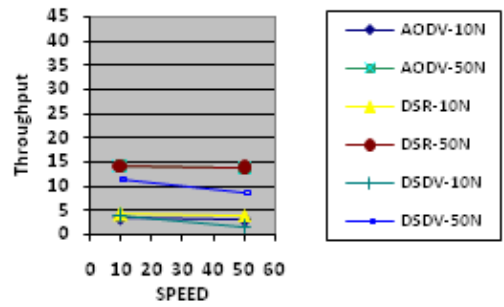

Fig 5(b)

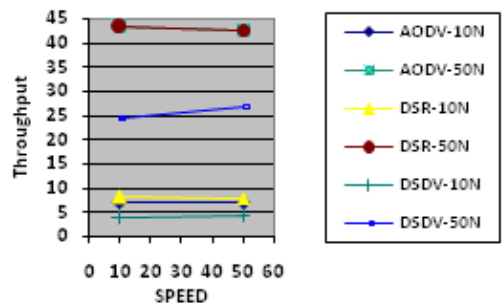

Fig 5(c)

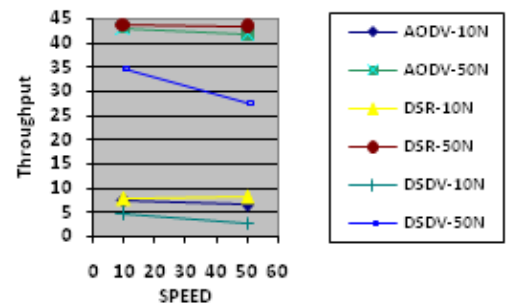

Fig 5(d)

# 7. CONCLUSION

In the end we say that random walk mobility model and random way point mobility model both are actually same mobility models apart from the pause time which is zero in Random Walk Mobility Model. There motion, direction and angle of motion, speed, etc are similar to each other. Our results showed clearly that all protocols perform same under these models. But if we increase the pause time in Random Way point mobility model, it decreases the mobility and so as the path breakage which results in difference of performance.

# 8. REFERENCES

[1] C. Perkins, E. B. Royer and S. Das (2003), "Ad hoc On-Demand Distance Vector (AODV) Routing", *RFC 3561, IETF Network Working Group*, July.

[2] T. Clausen and P. Jacquet (2003), "Optimized Link State Routing Protocol (OLSR)", *RFC 3262, IETF Network Working Group*, October 2003.

[3] V. Nazari, K. Ziarati (2006), "Performance Comparison of Routing Protocols for Mobile Ad hoc Networks", *IEEE* 2006.

[4] Josh Broch, David A.Maltz, David B. Johnson Yih-Chen Hu and Jorjeta Jetcheva, "A Performance Comparison of Multihop Wireless Ad Hoc Network Routing Protocols", ACM MOBICOM 98, Dallas, Texas. pp 25-30, October 1998.

[5] Jochen Schiller "Mobile Communications", Addision Wesley Longman Pvt.Ltd, India. 2000.

[6] H. Ehsan and Z. A. Uzmi (2004), "Performance Comparison of Ad HocWireless Network Routing Protocols", *IEEE INMIC* 2004.

[7] F. Bertocchi, P. Bergamo, G. Mazzin (2003), "Performance Comparison of Routing Protocols for Ad hoc Networks", *IEEE GLOBECOM* 2003.

[8] M. Sanchez and P. Manzoni. A java-based ad hoc networks simulator. In *Proceedings of the SCS Western Multiconference Web-based Simulation Track*, Jan. 1999.

[9] T. Camp, J. Boleng, and V. Davies, A Survey of Mobility Models for Ad Hoc Network Research, in *Wireless Communication and Mobile Computing (WCMC): Special issue on Mobile Ad Hoc Networking: Research, Trends and Applications*, vol. 2, no. 5, pp. 483-502, 2002.

[10] S. R. Das, C. E. Perkins and E. M. Royer (2000), "Performance comparison of Two On-Demand Routing protocols for Ad hoc Networks", *In Proc. of INFOC OM* 2000, Tel Aviv, Israel, March 2000.

[11] R. Misra, C. R. Manda (2005)l, "Performance Comparison of AODV/DSR On-Demand Routing Protocols for Ad Hoc Networks in Constrained Situation", *IEEE ICPWC* 2005.

[12] CMU Monarch Group, "CMU Monarch extensions to the NS-2 simulator." Available from http://www.monarch.cs.cmu.edu/cmu-ns.html, 2006

[13] "The network simulator ns-2. http://www.isi.edu/nsnam/ns2,"

**Dr. Abdur Rashid Khan** has done PhD in Computer Science. His field of interest include Artificial Intelligence and Expert Systems, DSS, MIS, Software Engineering . Currently he is working as Associate Professor in ICIT department of Gomal University Pakistan.

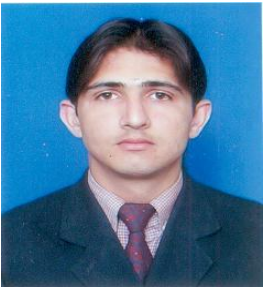**Mr. Muhammad Zaheer Aslam** has done BS in computer Science from Govt: Degree College NO.1 Dera Ismail Khan affiliated with Gomal University Dear Ismail Khan Pakistan. He has first division throughout his academic carrier. He has done his research on Mobile Adhoc Net (MANET). Currently, I am doing MSCS from gomal University Dera Ismail Khan Pakistan.